\documentclass{emulateapj}
\newcommand{\axp}{1E~1048.1$-$5937}
\newcommand{\axpu}{4U~0142$+$61}
\newcommand{\spz}{\textit{Spitzer}}

\begin{document}
\bibliographystyle{apj_noskip}

%\submitted{Submitted to ApJ Letters}

\title{Optical/Infrared Observations of the Anomalous X-ray Pulsar 1E~1048.1$-$5937 During Its 2007 X-Ray Flare}

\author{Zhongxiang Wang\altaffilmark{1}, Cees Bassa\altaffilmark{1}, Victoria M. Kaspi\altaffilmark{1}, Julia J. Bryant\altaffilmark{2}, and Nidia Morrell\altaffilmark{3}}
%\affil{ }

\altaffiltext{1}{Department of Physics,
McGill University, QC H3A 2T8, Canada}
%%\email{wangzx, vkaspi@physics.mcgill.ca}

\altaffiltext{2}{School of Physics, University of Sydney, NSW 2006, Australia}

\altaffiltext{3}{Las Campanas Observatory, Observatories of the Carnegie Institution of Washington, La Serena, Chile}

\begin{abstract}
We report on optical and infrared observations of the anomalous X-ray 
pulsar (AXP) \axp, made during its ongoing X-ray flare 
which started in 2007 March. We detected the source in the optical $I$ 
and near-infrared $K_s$ bands in two ground-based observations
and obtained deep flux upper limits from four observations, 
including one with the \textit{Spitzer Space Telescope} at 4.5 and 
8.0 $\mu$m. The detections indicate that the source was approximately 
1.3--1.6 magnitudes brighter than in 2003--2006,
when it was at the tail of a previous similar X-ray flare.
Similar related flux variations have been seen in two other AXPs during 
their X-ray outbursts, suggesting common behavior for large X-ray flux 
variation events in AXPs. 
The \spz\ flux limits are sufficiently deep that we can exclude 
mid-infrared emission similar to that from the AXP \axpu, 
which has been interpreted as
arising from a dust disk around the AXP. 
The optical/near-infrared emission from \axp\ probably has a magnetospheric
origin.  The similarity in the flux spectra of \axp\ and \axpu\ challenges 
the dust disk model proposed for the latter.

\end{abstract}

\keywords{stars: neutron --- pulsars: individual (1E 1048.1$-$5937) --- infrared: stars --- X-rays: stars}

\section{INTRODUCTION}

Supported by extensive observational studies over the past few years, 
it is generally believed that anomalous X-ray pulsars (AXPs) are 
magnetars---young neutron stars possessing ultra-high $\sim$10$^{14}$~G 
magnetic fields.  While AXPs are primarily known as X-ray sources, exhibiting
a variety of variability behavior related to their magnetar 
nature \citep{wt06,kas07}, we now know that 5 of 10 identified 
AXPs---4U~0142+61, \axp, 
1RXS~J170849$-$400910, XTE~J1810$-$197, and 1E~2259+586---are bright
at near-infrared (NIR) wavelengths (\citealt{wt06}; 
McGill AXP online catalog\footnote{\url{www.physics.mcgill.ca/pulsar/magnetar/main.html}}).  Among them, the AXP 4U 0142+61 is 
detected from optical to mid-infrared (MIR) wavelengths, indicating 
a spectral energy distribution (SED) that can be described by 
a two-component model: one a power-law spectrum over optical $V\/RI$ 
and NIR $J$ bands, presumably arising from the magnetosphere,
and one thermal blackbody-like over the 2.2--8 $\mu$m range 
\citep{wck06}, arising from a debris disk.
The discovery of such an optical/IR SED from the AXP was 
unexpected, and the two-component model remains controversial 
(e.g., \citealt{dv06d}). 
Ideally, in order to understand the optical/IR emission mechanism for 
AXPs and in particular determine whether the IR emission 
from 4U 0142+61 is unusual or generic among AXPs, SEDs of other AXPs 
must be observed.
However, most known AXPs are highly extincted in the optical range
and often extremely faint in the IR.

Among the known AXPs, \axp\ is peculiar in that
it has exhibited two long-term X-ray flares \citep{gk04, tam+08} which
have not been seen from other AXPs thus far.
Monitored with the \textit{Rossi X-ray Timing Explorer (RXTE)}, the first flare
was found to start in 2002 April and lasted approximately two years. 
At the beginning of the flare, the NIR counterpart
was discovered \citep{wc02}, and in observations more than a year later, 
the counterpart was found to be approximately 2 magnitudes dimmer in $JK_s$
\citep{dv05}, suggesting large flux variations associated with the flare. 
Such flaring events thus provide an opportunity to study optical/IR
emission from an AXP other than 4U 0142+61. 
In addition, during their X-ray outbursts, the AXPs 1E~2259+586 and
XTE J1810$-$197 both exhibited related NIR brightening. 
In the first source, its NIR and X-ray flux changes were correlated
\citep{tam+04} while in the latter, its NIR flux did not exactly follow
the decline of the X-ray flux (\citealt{rea+04}; \citealt{cam+07}).
Optical/IR observations of \axp\ during another X-ray 
flare would allow us to verify if
the source exhibits similar emission behavior.

For these reasons, after a second flare started on 2007 March 24 \citep{dib+07},
we made multi-wavelength observations of \axp\  with large ground-based
telescopes and the \textit{Spitzer Space Telescope}.
In this paper, we report on our optical/IR observations
during this on-going X-ray flare. 
%%Preliminary results of part of the observations have been reported \citep{wbk+07,wkb+07}.

\section{OBSERVATIONS AND DATA REDUCTION}    % Section 2
\label{sec:obs}

\subsection{Ground-Based Observations}

\subsubsection{NIR Imaging}

On 2007 April 04, we observed the \axp\ field in $K_s$ band
using Persson's Auxiliary Nasmyth Infrared Camera (PANIC; \citealt{mpm+04})
on the 6.5-m Baade Magellan Telescope at Las Campanas Observatory in Chile.
The detector is a Rockwell Hawaii 1024$\times$1024 HgCdTe array, having
a field of view (FOV) of 2\arcmin$\times$2\arcmin\  
and a pixel scale of 0.125\arcsec/pixel. The total on-source exposure time was 
15 min. During the exposure, the telescope was dithered in a
3$\times$3 grid with offsets of 10\arcsec\ to obtain a measurement of
the sky background. The observing conditions were good, with 0.5\arcsec\ 
seeing in $K_s$.

On 2007 May 08, we observed the target field in the same band
with the same telescope and camera. The total on-source time was 18 min, 
and the same observing
strategy was used. During the exposure, the conditions were not stable,
with $K_s$ seeing rapidly changing between 0.6\arcsec\ and 0.8\arcsec.

\subsubsection{Optical Imaging}

We observed the target field in the optical bands using
the 8.2-m Very Large Telescope (VLT; ANTU) at the European Southern 
Observatory (ESO)
and the 8-m Gemini South Telescope at the Gemini Observatory. Both telescopes
are located in Chile.

The VLT observation was made on 2007 May 07, the same night that the second
Magellan NIR image was obtained. The instrument used was the FOcal Reducer
and low dispersion Spectrograph (FORS2; \citealt{app+98}), which consists 
of two 2k$\times$4k
MIT CCDs and has a FOV of 6.8\arcmin$\times$6.8\arcmin.
%%The standard resolution collimator (COL\_SR) was used for the observation, 
The CCD detectors were 2$\times 2$ binned, having 
a pixel scale of 0.25\arcsec/pixel. We obtained 15 3-min $I$-band
images of the field, resulting in a 45 min total exposure. The telescope
was dithered in a 3$\times$5 grid, with offsets of 5\arcsec$\times$2.5\arcsec.
The conditions were excellent, with 0.5\arcsec\  seeing.

The Gemini imaging observations were made on 2007 June 24 and July 15.
The instrument was Gemini Multi-Object Spectrograph (GMOS; \citealt{hoo+04}). 
The detector array of GMOS consists of three 2048$\times$4608 EEV CCDs.
The pixel scale is 0.073\arcsec/pixel, while we used a detector 
binning of 2~pixels
for the observations. In the first observation, we obtained 8 $r'$ and 5 $i'$
images, with exposure times of 5 and 3 min, respectively. The telescope
was dithered for the exposures, with offsets of 10\arcsec.
The seeing was approximately 0.9\arcsec\ during the observations. In the latter observation,
we obtained 16 5-min $r'$ and 4 3-min $i'$ images, with the same 
observing strategy. The seeing was approximately
0.8\arcsec.

\subsubsection{Data Reduction}

We used the IRAF data analysis package for data reduction.
The images were bias-subtracted and flat-fielded.
In addition, because the Gemini GMOS detectors have significant fringing 
in $i'$ band, a fringe frame provided by the Gemini Observatory was used 
for subtraction of the fringes in our $i'$ images.
From each set of dithered images in one observation, 
a sky image was made by filtering out stars. The sky image was
subtracted from the set of images, and then the sky-subtracted images
were shifted and combined into one final image of the target field.
A summary of the images we obtained is given in Table~\ref{tab:obs}.

\subsection{\spz\  4.5/8.0 $\mu$m Imaging}

We also observed \axp\  on 2007 August 10 with the 
\textit{Spitzer Space Telescope}.
The imaging instrument used was the Infrared Array 
Camera (IRAC; \citealt{fha+04}). It operates in four channels 
at  3.6, 4.5, 5.8, and 8.0 $\mu$m, while two adjacent fields are 
simultaneously imaged in pairs (3.6 and 5.8 $\mu$m; 4.5 and 8.0 $\mu$m). 
We observed our target in the 4.5 (bandwidth 1.0 $\mu$m) and 
8.0 $\mu$m (bandwidth 2.9 $\mu$m) channels.
The detectors at the short and long wavelengths are InSb
and Si:As devices, respectively, with 256$\times$256 pixels and a plate
scale of 1\farcs2/pixel. The field of view (FOV) is 5\farcm2$\times$5\farcm2.
The frame time was 100 s, with 96.8 and 93.6 s effective exposure time
per frame for the 4.5 and 8.0 $\mu$m data, respectively. The total exposure
times in each observation were 53.2 min at 4.5 $\mu$m and 51.5 min at
8.0 $\mu$m.

The raw image data were processed through the IRAC data pipelines 
(version S16.1.0) at the {\em Spitzer} Science Center (SSC). 
In the Basic Calibrated Data (BCD) pipeline, standard imaging data
reductions, such as removal of the electronic bias, dark sky subtraction,
flat-fielding, and linearization, are performed and individual flux-calibrated 
BCD frames are produced.  In the post-BCD (PBCD) pipeline, radiation hits
in BCD images are detected and excluded, and BCD frames are then
combined into final PBCD mosaics.  The details of the data reduction
in the pipelines
can be found in the IRAC Data Handbook (version 3.0; \citealt{rsg+06}).

\section{RESULTS}
\label{sec:res}

\subsection{Ground-based Observations}

The counterpart to 1E 1048.1-5937 was detected in both NIR observations
made on 2007 April 04 and the optical observations made on 2007 May 07.
We performed PSF-fitting
photometry to measure the brightness of the source.  The nearby field
star X5 was used for flux calibration in the $K_s$ and $I$ bands \citep{dv05}.
The source's magnitudes were $K_s=19.9\pm0.1$ and $I=24.9\pm0.2$.
We did not detect the source in the other observations. Using the X5 star for
flux calibration \citep{wc02}, we derived 3$\sigma$ limiting magnitudes
for the Magellan $K_s$ and Gemini $r'$ and $i'$ images.
The results are given in Table~\ref{tab:obs}. 

\subsection{\spz/IRAC Observations}

We did not detect the source in the \spz\ IRAC 4.5 and 8.0 $\mu$m
images, and the derived 3$\sigma$ flux upper limits were
5.2 and 21.8 $\mu$Jy, respectively.  
The fluxes correspond to limiting magnitudes of 18.9 and 16.2 mag, 
for the zero magnitude fluxes of 179.7 and 64.1 Jy \citep{reach+05}
at the \spz/IRAC 4.5 and 8.0 $\mu$m bands, respectively. 
These results are also given in Table~\ref{tab:obs}.
The source region has been observed previously with IRAC at the
same wavelength bands in 2005, with no counterpart found either \citep{wkh07}. 
Comparing to the previous results, our 2007 upper limits are approximately
2 times deeper.

\section{DISCUSSION}
\label{sec:disc}

The optical/NIR counterpart of \axp\  was observed a few times during
2003--2006 after the discovery of the counterpart, and the source's 
brightness had been low
(e.g., $K_s\simeq 21.0$--21.5; see \citealt{tam+08} for details).
Compared to the flux measurements, particularly those obtained 
in 2003 April and June when the counterpart was detected in 
both the optical $I$ and NIR $JK_s$ bands (\citealt{dv05}; also see 
Table~\ref{tab:mea}), our results clearly indicate an optical/NIR brightening 
during the 2007 X-ray flare.  Indeed, assuming $A_V= 5.4$ mag, 
the unabsorbed optical/NIR--to--X-ray flux ratios for the 2003 and our 
measurements are nearly identical (see Figure~\ref{fig:comp}).
In this comparison, the unabsorbed 2--10 keV X-ray fluxes at two epochs  
given in Table~\ref{tab:mea} are used, and $A_V$ is
estimated from $N_{\rm H}=9.7\times 10^{21}$ cm$^{-2}$ (\citealt{tam+08}) 
by using the relation $A_V= N_{\rm H}/1.79\times 10^{21}$ cm$^{-2}$ 
\citep{ps95}.  Also we note that even though our
second $K_s$ measurement was an upper limit ($K_s>20.1$), it was 
only 0.1~mag smaller than that obtained from a detection two days later 
by \citet{isr+07}.
Thus in these two sets of observations, the optical/NIR 
and X-ray fluxes appeared to be correlated.

%\begin{figure}
\begin{center}
\includegraphics[scale=0.46]{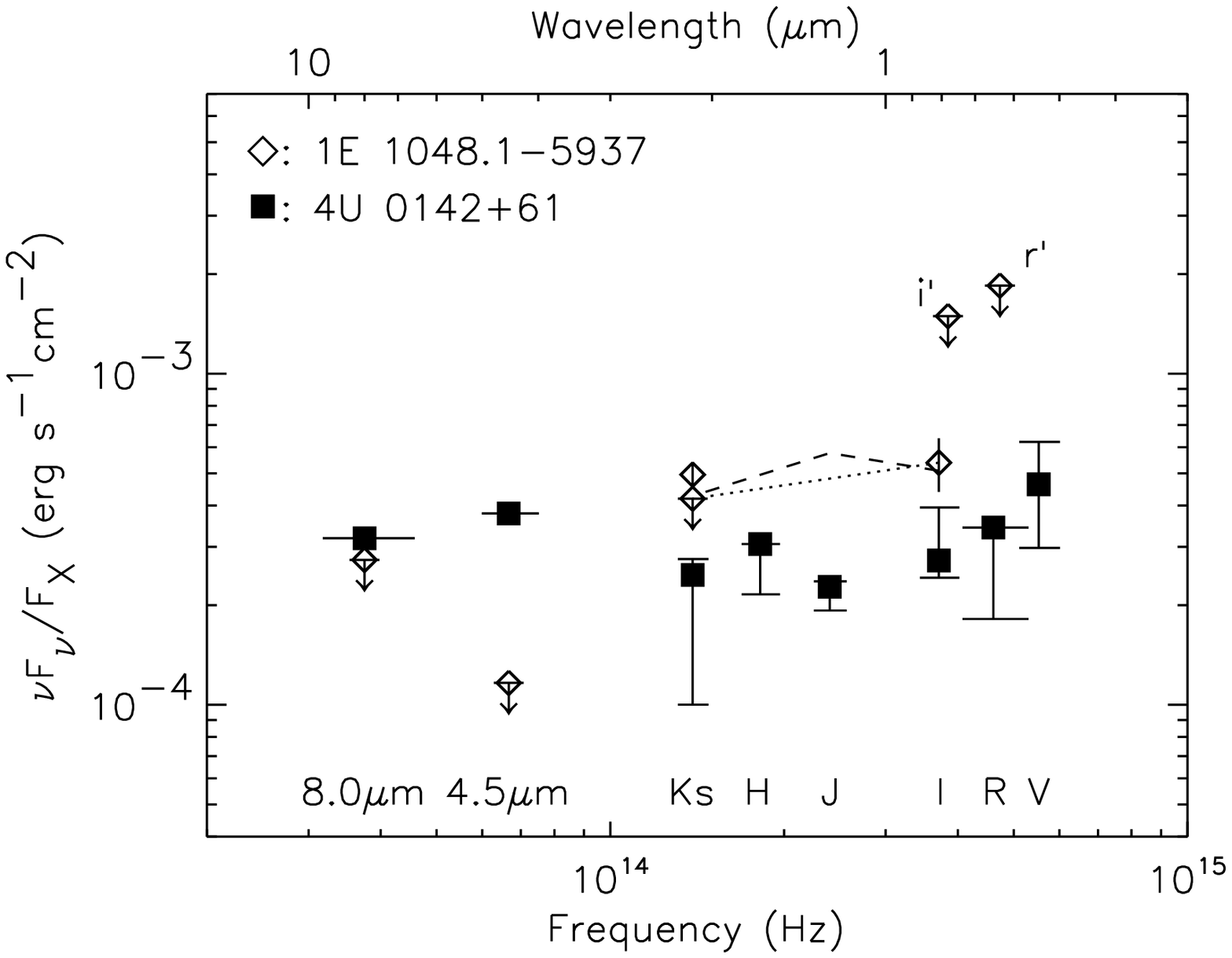}
%\plotone{f1.eps}
\figcaption{Unabsorbed optical/IR flux and flux upper limits (diamonds) of  
\axp\ measured in 2007 April--August, normalized by the unabsorbed 2--10 keV 
X-ray flux obtained on 2007 April 28 (see Table~\ref{tab:mea}). 
The dotted line indicates our two
nearly simultaneous observations in the $I$ and $K_s$ bands, and the dashed
line indicates the optical and NIR measurements obtained in 2003 April and June
\citep{dv05}.
For comparison, the unabsorbed optical/IR--to--X-ray flux ratios for 
\axpu\ are also shown (squares; \citealt{dv06d,wk08}), with the vertical 
and horizontal lengths of the bars indicating the flux
variation range and effective bandwidth at each band, respectively. 
$A_V=3.5$ \citep{dv06a} is used for dereddening.
The unabsorbed 2--10 keV X-ray flux used for \axpu\ was 6.6$\times 10^{-11}$ 
ergs s$^{-1}$ cm$^{-2}$ \citep{gon+07}.
\label{fig:comp}
}
\end{center}
%\end{figure}

However, as shown by \citet{tam+08}, the observed NIR brightening at 
the beginning of the 2002 X-ray flare preceded the pulsed X-ray flux peak 
by $\sim$2 months. If we assume that the relation between the pulsed 
flux $F_{\rm p}$ and total flux $F_{\rm t}$, 
$F_{\rm p} \propto F_{\rm t}^{0.54}$ \citep{tam+08}, is generally true 
for this AXP, the 2--10 keV unabsorbed total flux derived at the NIR 
observation time would have been $\simeq$$10^{-11}$ ergs s$^{-1}$ cm$^{-2}$, 
3.6 times lower than the 2007 peak flux.  As a result, the 2002 
NIR--to--X-ray flux ratios would be approximately 6 times higher than 
that shown in Figure~\ref{fig:comp} (note $K_s=19.4$ in the 
2002 detection). This would indicate that the NIR and X-ray fluxes were 
not correlated all the time. 
We note that since the $F_{\rm p}$ and $F_{\rm t}$ relation is mainly
based on the total flux measurements obtained during the 2007 flare and there 
were not such measurements for the 2002 flare, it is possible
that the total flux was much larger at the start of the 2002 flare, and not 
reflected in the \textit{RXTE} pulsed flux measurements.
In any case, we confirm that in this AXP, the large optical/NIR flux variations 
were related to the X-ray flares. Similar related NIR brightening has been 
seen in 1E~2259$+$586 and XTE~J1810$-$197 during their X-ray outbursts
\citep{tam+04,cam+07}, suggesting common behavior for emission from AXPs.
However for \axpu, while its X-ray flux
has been relatively stable \citep{gon+07}, its optical and NIR flux was found to
have large, rapid variations \citep{dv06d}. This limits the 
related NIR/X-ray behavior to only large X-ray flux variation events.

It is instructive to compare \axp\ to \axpu, since the latter has
well measured fluxes from the optical to MIR.
In Figure~\ref{fig:comp}, we compare the optical/IR fluxes and flux upper 
limits of \axp\ to those of \axpu, with each set of data points 
normalized by the approximately contemporaneous
unabsorbed 2--10 keV X-ray flux of each source. The optical $V\/RI$ and 
NIR $JHK_s$ fluxes 
for \axpu\ are from Durant \& van Kerkwijk (2006b; because 
of relatively large
flux variations from the source, we show the range of the flux found
in each band in the figure), the two MIR data points at 4.5 
and 8.0 $\mu$m are the 2005 flux measurements in Wang \& Kaspi (2008), 
and the unabsorbed 2--10 keV X-ray flux
used is 6.6$\times 10^{-11}$ ergs s$^{-1}$ cm$^{-2}$, obtained
on 2004 July 24 \citep{gon+07}. As can be seen from the figure,
while the flux ratios of \axp\ are roughly two times larger than 
those of \axpu, the deep limit at 4.5 $\mu$m
from our \spz\ IRAC observation is approximately 3 times lower than
the detection of \axpu. This strongly suggests that
our target does not have similar MIR emission.
This difference in the MIR is not very sensitive to 
uncertainties on the reddening to the two sources, since MIR emission 
is only weakly extincted by the interstellar medium.
The X-ray flux from \axpu\ has been stable \citep{gon+07}.
A possible explanation could be the non-simultaneous X-ray flux of \axp\ 
we use for the comparison. However, \textit{RXTE} 
X-ray monitoring observations of the source have indicated that
the pulsed flux, which is found to trace the total 
flux \citep{tam+08}, has remained stable and high during our observations 
(Dib et al. 2008, in preparation).  In order to raise the 4.5 $\mu$m
upper limit to the flux ratio of \axpu, the X-ray flux would have
had to have been
as low as 11$\times 10^{-12}$ ergs s$^{-1}$ cm$^{-2}$, only 40\% larger
than the average quiescent flux (7.7$\times 10^{-12}$ ergs s$^{-1}$ cm$^{-2}$)
of the source and approximately one third of those obtained 
in 2007 April \citep{tam+08}.

The lack of MIR emission similar to that of \axpu\  from \axp\ seems robust. 
In the dust disk model proposed for \axpu\ \citep{wck06},
the MIR flux, arising from X-ray irradiation of the disk, is 
proportional to the X-ray flux of the pulsar. 
Therefore, even though our deep 4.5 $\mu$m limit was obtained during 
the X-ray flare, we can exclude the existence of a similar disk around \axp.
However, since a disk could be further away from the central source,
deep observations at longer wavelengths are needed in order to exclude
the existence of a disk more conclusively.
The MIR non-detection may suggest that debris disks are not 
commonly found around AXPs and thus the putative disk in 
\axpu\ is unique. 
Two other AXPs have also been observed by \spz\  with no detection;
however, the derived upper limits are far above the MID--to--X-ray flux ratio 
of \axpu\ \citep{wkh07}.

The optical and NIR emission from \axp\ probably has a magnetospheric origin,
given that the flux spectrum is similar to that of \axpu\ 
(\citealt{wc02,dv05}). For the latter source, its optical 
emission is known to be pulsed at the spin period and has a pulsed fraction 
much higher than that in X-rays, excluding a 
disk origin \citep{km02,dhi+05}. 
Details of how optical and NIR emission is produced in the magnetosphere of a 
magnetar are not known, though possible 
radiation mechanisms have been suggested (e.g., \citealt{bt07}). 
We note that in the disk model for \axpu\ \citep{wck06}, the $K$-band flux 
primarily arises from the disk, not from the magnetosphere. 
Therefore, the similarity in the $K$-band--to--X-ray flux ratios for 
the AXPs \citep{dv05}, including \axp\ as confirmed by
our measurements, challenges the disk model. 
The $K$-band emission thus should originate from the magnetosphere, which 
may be supported by the fact that in the absence of X-ray variability,
strong $K$-band variability is seen in \axpu\ \citep{dv06d}. 

\acknowledgements
We thank ESO, the Gemini Observatory, and SSC for granting us the observations.
We thank F. Patat at the Users Support Department of ESO, M. Bergmann at
Gemini South, and N. Silbermann from SSC for helping with the observations.
We also thank R. Dib and C. Tam for sharing their unpublished results.
The Gemini data were taken under the program DD-2007A-DD-10-1.
The Gemini Observatory is operated by the Association of Universities 
for Research in Astronomy, Inc., under a cooperative agreement with 
the NSF on behalf of the Gemini partnership: the National Science Foundation 
(United States), the Science and Technology Facilities Council 
(United Kingdom), the National Research Council (Canada), CONICYT (Chile), 
the Australian Research Council (Australia), CNPq (Brazil), and CONICET 
(Argentina). 
This research was supported by NSERC via a Discovery Grant
and by the FQRNT and CIFAR.  VMK holds a Canada Research Chair and
the Lorne Trottier Chair in Astrophysics \& Cosmology, and is a
R. Howard Webster Foundation Fellow of CIFAR.

\bibliographystyle{apj}
%%\bibliography{axp}

\clearpage

\begin{deluxetable}{l c c c c c c c}
\tabletypesize{\scriptsize}
\tablewidth{0pt}
\tablecaption{Optical/IR Observations of \axp\label{tab:obs}}
\tablehead{
\colhead{Observation Start} & \colhead{MJD} & \colhead{Telescope/Instrument} & \colhead{Filter}
& \colhead{Exposure}  & \colhead{Seeing\tablenotemark{a}} & \colhead{Magnitude\tablenotemark{b}} & \colhead{ $\nu F_{\nu}$\tablenotemark{c}/10$^{-14}$ } \\
\colhead{Time (UTC)}            & \colhead{} & \colhead{}  & \colhead{} 
& \colhead{(min)} & \colhead{(arcsec)} & \colhead{} & \colhead{ (ergs s$^{-1}$ cm$^{-2}$) }   }
\startdata
2007-04-04 00:37 & 54194.0 &Magellan/PANIC  & $K_s$ & 15 & 0.54 & 19.9$\pm$0.1  & 1.8 \\ 
2007-05-07 23:24 & 54228.0 &VLT/FORS2	   & $I$   & 45 & 0.55 & 24.9$\pm$0.2   & 1.9 \\
2007-05-08 03:15 & 54228.1 &Magellan/PANIC  & $K_s$ & 18 & 0.66 & $>$20.1       & $<$1.5 \\ 
2007-06-24 01:18 & 54275.1 &Gemini-S/GMOS   & $r'$  & 40 & 1.0  & $>$24.8 & \nodata \\
	         &	   &	   & $i'$  & 15 & 0.88 & $>$24.1 & \nodata \\
2007-07-15 00:06 & 54296.0 &Gemini-S/GMOS   & $r'$  &	80 & 0.92 &  $>$25.6    & $<$6.6 \\
                 &         &       & $i'$  &	12 & 0.74 & $>$24.5 & $<$5.4 \\
2007-08-10 10:04 & 54322.4 & Spitzer/IRAC    & 4.5 $\mu$m & 53.2 & \nodata & $>18.9$ & $<$0.42 \\
	         & 	   &                 & 8.0 $\mu$m & 51.5 & \nodata & $>16.2$ & $<$0.99 \\
\enddata
\tablenotetext{a}{Full width at half maximum.}
\tablenotetext{b}{The upper limits are 3$\sigma$ limiting magnitudes.}
%%\tablenotetext{c}{The zero magnitude flux for the \spz/IRAC 4.5 and 8.0 $\mu$m bands are 179.7 and 64.1 Jy, respectively \citep{reach+05}.}
\tablenotetext{c}{Flux is dereddened with $A_V= 5.4$, using the reddening laws 
for the optical and NIR bands given in \citet{sfd98} and for the \spz/IRAC
data given in \citet{imb+05}.}
\end{deluxetable}

%%\clearpage
\begin{deluxetable}{l c c c c }
\tabletypesize{\scriptsize}
\tablewidth{8cm}
\tablecaption{Flux measurements of \axp\  used in the comparison.
\label{tab:mea} 
}
\tablehead{
\colhead{Epoch} & \multicolumn{2}{c}{Optical/NIR} & \colhead{X-ray\tablenotemark{a}} & \colhead{Refs} \\
   & \colhead{Magnitude} & \colhead{ $\nu F_{\nu}$\tablenotemark{b} } &  &  
}
\startdata
2003 April 24 & $J=23.4$   & 6.6     & \nodata & 1 \\
2003 June 6   & $I=26.2$   & 5.9     & \nodata & 1 \\
2003 June 7   & $K_s=21.5$ & 4.9     & \nodata & 1 \\
2003 July 16  & \nodata    & \nodata & 11.5    & 2 \\[1ex]
2007 April 28 & \nodata    & \nodata & 36.1    & 2 \\
\enddata
\tablerefs{
(1) \citet{dv05};
(2) \citet{tam+08}.}
\tablenotetext{a}{X-ray flux is phase-averaged, unabsorbed, in the energy 
range of 2--10 keV, and in units of 10$^{-12}$ ergs s$^{-1}$ cm$^{-2}$.} 
\tablenotetext{b}{Dereddened flux (assuming $A_V=5.4$; in units of 10$^{-15}$ ergs s$^{-1}$ cm$^{-2}$) obtained at the tail of the first X-ray flare.}
\end{deluxetable}

\end{document}